\documentclass[11pt]{article}
\begin{document}

\begin{center}
{\bf \large{Electrodynamics at non-zero temperature, chemical 
potential, and Bose condensate}} \\ \vspace{0.5cm}
{\it Alexander D. Dolgov \footnote{dolgov@fe.infn.it}$^{a,b,c}$, 
Angela Lepidi \footnote{lepidi@fe.infn.it}$^{a,b}$,
 and Gabriella Piccinelli \footnote{gabriela@astroscu.unam.mx}$^{d,a}$}
\\\vspace{0.3cm}
$^a$ Istituto Nazionale di Fisica Nucleare, Sezione di Ferrara, 
I-44100 Ferrara, Italy \\
$^b$ Dipartimento di Fisica, Universit\`a degli Studi di Ferrara, 
I-44100 Ferrara, Italy \\
$^c$ Institute of Theoretical and Experimental Physics, 113259 Moscow, Russia \\
$^d$ Centro Tecnol\'ogico, FES Arag\'on, Universidad Nacional Aut\'onoma de M\'exico, 
Avenida Rancho Seco S/N, Bosques de Arag\'on, Nezahualc\'oyotl, 
Estado de M\'exico 57130, Mexico

\begin{abstract}
Electrodynamics of charged scalar bosons and spin $1/2$ fermions is studied 
at non-zero temperature, chemical potentials, and possible Bose condensate of 
the charged scalars. Debye screening length, plasma frequency, and the photon 
dispersion relation are calculated. It is found that in presence of the 
condensate the time-time component of the photon polarization operator
in the first order in electric charge squared
acquires infrared singular parts proportional 
to inverse powers of the spatial photon momentum $k$.
\end{abstract}
\end{center}

\section{Introduction}

Gauge theories in medium with non-zero temperature $T$ and chemical potential 
$\mu$ are of interest in plasma physics, in particular for quark-gluon 
plasma \cite{Arnold:2007pg,plasma-book}, astrophysics~\cite{plasma-book,astro} 
and cosmology~\cite{Linde:1978px,cosmo}. While in heavy ion
collisions, where quark-gluon plasma might presumably form, and in stellar environment chemical
potentials are typically large, $\mu \geq T$, in cosmology it is usually assumed that $\mu \ll T$. 
This assertion is based on the tiny value of the observed baryon asymmetry, 
$\beta =(N_B - N_{\bar B}) /N_\gamma \approx 6\cdot 10^{-10}$ and the bounds on the 
cosmological lepton
asymmetry \cite{Dolgov:2002ab}. However, it is not excluded that in the early universe 
some charge asymmetries, including the lepton one, might be large~\cite{ad-ft}. 

Bearing in mind possible applications to cosmology and, maybe, to other
mentioned above fields, 
we consider in this paper a simple case 
of an Abelian gauge theory, namely electrodynamics and focus on the photon propagation 
in hot degenerate plasma consisting of charged scalar bosons 
with mass $m_B$ and fermions with mass $m_F$.
This problem is interesting by itself from theoretical point of view and 
may be useful for understanding phenomena in heavy nuclei or neutron and quark stars 
as well as in the early universe.
A lot of work has been done in gauge theories at non-zero $T$, see e.g. books~\cite{Kapusta:1989tk}
and references therein. 
The usual spinor electrodynamics (SpQED) at $T\neq 0$ was studied ages ago in classical plasma physics,
see e.g.~\cite{landau-9}, and later many works were dedicated to the investigation of SpQED 
in the frameworks of relativistic quantum field theory. 
Much less attention was devoted in earlier years to the scalar quantum electrodynamics (SQED) 
in medium by an evident reason that no fundamental charged particles were known. However subsequent
theoretical discoveries of supersymmetry (SUSY) and spontaneous symmetry breaking (SSB), 
and possible existence of fundamental scalars in some extension of the 
Minimal Standard Model revived interest
to SQED at non-zero temperatures, for a review see e.g.~\cite{Kraemmer:1994az}. 
 
Usually SQED was studied at non-zero $T$ but with vanishing chemical potential of bosons,
because the medium was assumed to be electrically neutral.
There are some works that invoke a chemical potential for bosons and also deal 
with Bose condensation. 
They are typically interested in the relationship between spontaneous symmetry breaking 
and Bose-Einstein condensation, as in \cite{Kapusta:1981aa,Benson:1991nj} 
and do not consider aspects regarding photon propagation in the medium.
On another hand, the propagation of a heavy quark through a quark-gluon plasma 
in presence of a quark chemical potential is considered in \cite{Vija:1994is}.

In our work we do not confine ourselves to the restriction $\mu =0$ and moreover we 
consider, in particular, 
the maximum allowed value of the chemical potential of scalar bosons which is
equal to their mass, $\mu_B = m_B$, and at which they may Bose condense. 
The SQED with Bose condensate of charged scalar bosons was not 
considered in the previous literature, except for the
two recent papers~\cite{Gabadadze:2007si,gaba-2},
where somewhat different issues were studied. In the 
above mentioned papers and in this one
the medium was assumed to be electrically neutral because of 
the assumed compensation of electric charge
density of bosons by the opposite charge density of fermions. In ref.~\cite{Gabadadze:2007si}
a  conductive spherical object of a finite size, e.g. a star, was considered and the 
neutralization  of the interior was achieved by forcing an excessive charge to the surface. 
In ref.~\cite{gaba-2} as well as here the charge neutrality of the medium was assumed
{\it ab ovo}  i.e. by 
the initial condition that the total (conserved) electric charge densities of bosons 
and fermions are equal by magnitude and opposite by sign. The restriction of the overall
charge neutrality is not necessary and cosmology with non-vanishing electric charge density
might exist (see below).

A similar situation to that discussed in the present work
was studied in ref.~\cite{Linde:1976kh} where a medium with
charged scalars and fermions, each with non-zero charge density, was considered. 
The paper was dedicated to different problems but not to the calculation of the 
polarization tensor of photon in such a medium, which is a primary aim of the
present work. The charged boson state in ref.~\cite{Linde:1976kh} was described
in a different from ours gauge which is not convenient for the 
perturbation theory which is used in this paper. 
Since the expression for the polarization operator is gauge invariant in QED,
it should be the same in any gauge.
Nevertheless we have explicitly verified that our result for the 
photon polarization operator remains the
same in the gauge of ref.~\cite{Linde:1976kh} as well.

There are two standard approaches to field theory at nonzero temperature and chemical potentials:
either imaginary time method (Matsubara formalism) or real time method 
(Schwinger-Keldysh formalism). The former
is applicable only to the case of systems in thermal equilibrium, while the latter is valid for
any state of the medium.  For a review of these methods, see e.g. ref.~\cite{high-T-rev}
or books~\cite{Kapusta:1989tk}. Below we
calculate the photon Green's function for arbitrary medium in a physically
transparent and simple way taking the expectation
values of relevant operators not only over vacuum but over any state of the system, either
it is a collection of particles with arbitrary occupation numbers or a coherent field state (the
latter is possible for bosons only). 
We calculate the photon polarization operator in the medium and derive from it the
electrostatic (Debye) screening length, the plasma frequency and more generally the photon 
dispersion relation. In all previously studied cases our results coincide with the 
known ones.

When setting the boson chemical potential equal to its limiting value, $\mu_B = m_B$, 
we found a surprising equation of motion for the photon:
the correction to the time-time component of the photon polarization operator 
at zero frequency contains in addition 
to the free term $G^{-1}_{00} = k^2$ and the well known momentum independent term, which 
is inversely proportional to the Debye screening length squared if $\mu <m_B$,
two additional terms, one of which appears when the 
bosonic chemical potential reaches its maximum value, $\mu = m_B$, and 
is inversely proportional to the photon momentum, i.e. $1/k$, and another term which appears
when bosons condense and which is inversely proportional to the photon momentum squared,
i.e. $1/k^2$. In other words the Maxwell equation for the time component of the potential
at zero frequency (stationary case) and in the low momentum limit is modified as:
\begin{eqnarray}
- [k^2 + e^2(m_0^2 + m_1^3/k + m_2^4/ k^2)] A_0 = J_0,
\label{max-eq}
\end{eqnarray}
where $m_j$ are some constant parameters expressed through the characteristics of the medium and
$J_0$ is the electric charge density of the medium, see below eq. (\ref{current}).
As is mentioned above, we assumed that $J_0 = 0$ to avoid ``infrared''
problems, though this restriction is not necessary, especially in cosmology. 
The $1/k^2$ term was also found by G. Gabadadze (private communication, work in progress). We
thank him for sending the draft of his work prior to publication. In the paper 
by Gabadadze a zero temperature 
case is considered and because of that $1/k$ term does not appear. As shown in the present work, 
the $1/k$ term leads to a power law decrease of the screened potential, eq.~(\ref{power-U}).

To study the
screening phenomenon we take, as usually, $J_0$ to be a point-like 
test charge,  $J_0 \sim q\delta^{(3)} ({\bf k}) $.
Evidently the Debye screening mass in this case is not equal, as usually, to    
$e m_0$ but to more complicated expression involving $m_1$ and $m_2$ and possessing a non-zero 
imaginary part which leads to an oscillating behavior of screening superimposed on the exponential
decay, see eq.~(\ref{U-j}) . 
Similar oscillating behavior of the 
screening was also observed in ref.~\cite{gaba-2}. The oscillations of the screened electrostatic
potential was found long ago in non-relativistic fermionic plasma, see e.g. book~\cite{fetter-walecka}.
They are known as Friedel oscillations.
More recently a similar phenomenon was investigated in ref.~\cite{kapusta-88} for 
relativistic QED and QCD \footnote{We thank G. Gabadadze for indicating these references.}.
The oscillating behavior due to fermions appears only in higher order in the coupling
constant, $\alpha = e^2/4\pi$, while oscillations due to Bose condensate appear in the lowest
order in the coupling. 

The paper is organized as follows: in Sec. \ref{sec_SQED_calc} we introduce the photon 
polarization tensor $\Pi^{\mu\nu}$ and the techniques we used to work it out. 
In Sec \ref{sec_coll_phen} we use the appropriate components of $\Pi^{\mu\nu}$ to calculate the 
Debye screening mass $m_{D}$ and the plasma frequency $\omega_p$. 
Some limiting cases, which do not include condensate, are analyzed to check compatibility with 
previous results found in literature. 
In Sec. \ref{Calcs-in-other-gauge} we calculate the Debye mass in an explicitely fixed gauge.
Finally in Sec. \ref{sec_concl} our conclusions are presented.

\section{Perturbative calculations of the photon polarization operator}
\label{sec_SQED_calc}

The Lagrangian of interacting electromagnetic field and charged scalar and fermion fields 
with masses $m_B$ and $m_F$ respectively and with opposite electric charges $\pm e$ has the form:
	\begin{eqnarray}
	\label{L_SQED}
	\mathcal{L} = -\frac{1}{4} F_{\mu\nu} F^{\mu\nu} - m_B^2 |\phi|^2
	+ |(\partial_{\mu} + i\,e A_{\mu}) \phi |^2
	+ \bar \psi (i \partial \!\!\! / - e A \!\!\! /  - m_F) \psi.
	\end{eqnarray}
The Lagrangian is symmetric under the gauge transformations: 
	\begin{eqnarray}
	\label{gauge-trnsfr}
	\phi(x) \rightarrow \exp [ i e \alpha(x)] \: \phi(x),
	\hspace{1cm}
	\psi(x) \rightarrow \exp [ i e \beta(x)] \: \psi(x),
\hspace{1cm}
	A_\mu (x) \rightarrow A_\mu(x) - \partial_\mu(\alpha+\beta),	
\end{eqnarray}
This implies an existence of $2$ conserved currents and charges, which we can choose as the 
scalar and the fermion number. 
The Lagrangian (\ref{L_SQED}) leads to the following
equations of motion for the involved fields: 
	\begin{eqnarray}
	\label{Psi_EOM}
	 (i \partial \!\!\! /  - m) \psi (x) = e A \!\!\! / \psi (x)
	\end{eqnarray}
	\begin{eqnarray}
	\label{Phi_EOM}
	(\partial_{\mu} \partial^ {\mu} + m^2) \phi (x) &=& \mathcal{J}_\phi (x)
	\end{eqnarray}
	\begin{eqnarray}
	\label{phot_EOM}
	\partial_{\nu} F^{\mu\nu} (x) &=& \mathcal{J}^\mu (x)
	\end{eqnarray}
where the currents $\mathcal{J}$ are defined as 
	\begin{eqnarray}
	\label{phi_current}
	\mathcal{J}_\phi (x) &=&
	-  i\,e \bigg[  \partial_{\mu} A^{\mu} (x) + 2 A_{\mu} (x) \partial^{\mu} \bigg] \phi (x)
	+ e^2 A^{\mu} (x) A_{\mu} (x) \phi (x) 
	\end{eqnarray}
%
%
	\begin{eqnarray}
	\label{EOM_current}
	\mathcal{J}^\mu (x)
	&=& - i\,e \bigg[(\phi^{\dag} (x)\partial^{\mu} \phi(x) ) - 
(\partial^{\mu} \phi^{\dag}(x)) \phi(x) \bigg]
	+ 2 e^2 A^{\mu} (x)|\phi(x)|^2 
	- e \bar \psi (x) \gamma^\mu \psi (x).
	\end{eqnarray}
Here $F_{\mu\nu} = \partial_\mu A_\nu - \partial_\nu A_\mu$ and
$\mathcal{J}^\mu$ (\ref{EOM_current}) is the total electromagnetic current 
of bosons and fermions in the coordinate space.

The key quantity which determines the photon propagation in plasma
is the photon polarization tensor $\Pi^{\mu\nu}$ which we will calculate perturbatively. 
When doing this kind of calculations for massless fields, infrared singularities may
arise and to regularize them one should use the resummation techniques - 
see e.g. \cite{Kapusta:1989tk} and references therein. Nevertheless it is safe to use the 
standard perturbative solution when the scalar and 
fermion field masses are not negligible as in the case we are considering. Moreover, the
infrared singularities in Abelian theories are much milder than those in non-Abelian ones
where the correspondence between the order of the perturbative series and the power of 
the coupling constant $e$ is lost, see e.g. discussion in 
ref.~\cite{Linde:1978px,linde-book}. Here we consider only Abelian QED and since it is 
not infrared dangerous, we neglect the resummation.

We formally solve operator equations (\ref{Psi_EOM}) and (\ref{Phi_EOM}) as:
	\begin{eqnarray}
	\label{field_1_ord_exp}
	\phi(x) &=& \phi_0 (x) + \int d^4y \, G_B(x-y) \mathcal{J}_{\phi} (y)
	\cr\cr
	\psi(x) &=& \psi_0 (x) + \int d^4y \, G_F(x-y) e A\!\!\! / (y) \psi(y)
	\end{eqnarray}
where the zeroth order fields satisfies the free equations of motion:
	\begin{eqnarray}
	\label{phi_0_ord}
	(\partial_{\mu} \partial^ {\mu} + m_B^2) \phi_0 (x) = 0 ,
	\hspace{1cm} 
	( i \partial \!\!\! / - m_F ) \psi_0 (x) = 0 
	\end{eqnarray}
and are quantized in the usual way
	\begin{eqnarray}
	\label{phi-0-x}
	\phi_0 (x) &=& \int \frac{d^3q}{\sqrt{(2\pi)^3 2 E}}
	\left[ a({\bf q}) \exp^{-i q x} + b^\dag ({\bf q}) \exp^{iqx} \right]
	\cr\cr
	\psi_0 (x) &=& 
	\int \frac{d^3q}{\sqrt{(2\pi)^3}} \sqrt{\frac{m_F}{E}}
	\left[ c_r({\bf q}) u_r({\bf q}) \exp^{-i q x} + d_r^\dag 
({\bf q}) v_r ({\bf q}) \exp^{iqx}
	\right].
	\end{eqnarray}
In eq. (\ref{phi-0-x}) $a^{(\dag)}$, $b^{(\dag)}$, $c^{(\dag)}$ and $d^{(\dag)}$ are the 
annihilation (creation) operators for scalar and spinor particles and antiparticles.
The Green functions in eqs. (\ref{field_1_ord_exp}) are the usual Feynman Green functions 
having the integral representation:
	\begin{eqnarray}
	\label{Green_funct}
	G_{B,F} (x-y) = \int \frac{d^4 k}{(2 \pi)^4}  \exp^{-i k (x-y)} G_{B,F}(k) 
	 \end{eqnarray}	
where
\begin{eqnarray}
G_B(k) =  \frac{1}{k^2 - m_B^2 + i\epsilon}, \,\,\,{\rm and}\,\,\,
	G_F (k) = \frac{k \!\!\! / + m_F}{k^2 - m_F^2 +i \epsilon}.
	\end{eqnarray}
%
Now we can substitute eqs. (\ref{field_1_ord_exp}) into eq. (\ref{phot_EOM}) with 
the currents 
given by eqs. (\ref{phi_current}) and (\ref{EOM_current}). 
For the calculations up to the second order in the coupling constant $e$, i.e. up 
to $e^2$, it is 
sufficient to include into $\mathcal{J}_\phi$ in eq. (\ref{phi_current})
only terms of the first order in $e$, that is
	\begin{eqnarray}
	\label{J-phi-x}
	\mathcal{J}_\phi (x) &=&
	+ i\,e \bigg[  \partial_{\mu} A^{\mu} (x) + 2 A_{\mu} (x) \partial^{\mu} \bigg] 
\phi_0 (x).
	\end{eqnarray}
As a result we obtain
	\begin{eqnarray}
	\label{partial_drv_Fmunu}
	\partial_\nu F^{\mu\nu} (x) &=&
	-i\,e \bigg[(\phi_0^{\dag} (x)\partial^{\mu} \phi_0(x) ) 
	- (\partial^{\mu} \phi_0^{\dag}(x)) \phi_0(x) \bigg]
	- e \bar \psi_0 (x) \gamma^\mu \psi_0 (x) 
	\cr\cr
	&-& i e \, \phi_0^\dag (x) \partial^{\mu} \left[ \int d^4y \, G_B(x-y) 
	\mathcal{J}_{\phi_0} (y) \right]
	- i e \left[ \int d^4y \, G_B(x-y) \mathcal{J}_{\phi_0} (y) \right]^\dag 
	\partial^{\mu}  \phi_0 (x)
	\cr\cr 
	&+& i e \, \partial^{\mu} \phi_0^{\dag}(x) \left[ \int d^4y \, G_B(x-y) 
	\mathcal{J}_{\phi_0} (y) \right]
	+ i e \, \partial^{\mu} \left[  \int d^4y \, G_B(x-y) \mathcal{J}_{\phi_0} (y) \right] ^\dag
	\phi_0(x)
	\cr\cr
	&-& e \bar \psi_0 (x) \gamma^\mu \int d^4y \, G_F(x-y) e A\!\!\! / (y) \psi(y)
	- e  \left[ \int d^4y \, \bar \psi_0 (y) A\!\!\! / (y) \, G_F^*(x-y) \right] \gamma^\mu \psi_0(x)
	\cr\cr
	&+& 2 e^2 A^{\mu} (x)|\phi_0(x)|^2 .
	\end{eqnarray}
To derive the Maxwell equations with the account of the impact of medium on 
the photon propagator 
we have to average operators $\phi$ and $\psi$ over the medium. 
The first term in eq. (\ref{partial_drv_Fmunu}), linear in $e$, is non-zero if 
the medium is either 
electrically charged or possesses an electric current.
The products of creation-annihilation operators averaged over the medium have the 
standard form:
	\begin{eqnarray}
	\label{thermal_averages}
	\langle a^\dag({\bf q}) a({\bf q}') \rangle &=& 
f_B (E_q) \delta^{(3)} ({\bf q} - {\bf q}'),
	\hspace{1.5cm}
	\langle a({\bf q}) a^\dag({\bf q}') \rangle = [1 + f_B (E_p)] \delta^{(3)} 
({\bf q} - {\bf q}'),
	\cr\cr
	\langle c^\dag({\bf q}) c({\bf q}') \rangle &=& f_F (E_p) \delta^{(3)} 
({\bf q} - {\bf q}'),
	\hspace{1.5cm}
	\langle c({\bf q}) c^\dag({\bf q}') \rangle = [1 - f_F (E_p)] \delta^{(3)} 
({\bf q} - {\bf q}'),
	\end{eqnarray}
where $f_{F,B}(E_q)$ is the energy dependent fermion/boson distribution function, 
which may be 
arbitrary since we assumed only that the medium is homogeneous and isotropic.
We also assumed, as it is usually done, that non-diagonal matrix elements of
creation-annihilation operators vanish on the average due to decoherence.
For the vacuum case $f_{F,B}(E) = 0$ and we obtain the usual vacuum average values of 
$a a^\dag$ and $a^\dag a$, which  from now on will be neglected because we are 
interested only in the matter effects.
As a result we obtain linear but non-local equation for electromagnetic field $A_\mu (x)$.
It is convenient to perform the Fourier transform
	\begin{eqnarray}
	\label{A-mu-k}
	A^\mu (k) = \int \frac{d^4 x}{(2\pi)^3} \exp^{-ikx} A^{\mu} (x). 
	\end{eqnarray}
Finally we find that the field $A^\mu (k)$ satisfies the equation
	\begin{eqnarray}
	\label{Phot_EOM_Pi_munu}
	\left[ k^\rho k_\rho g^{\mu\nu} - k^\mu k^\nu + \Pi^{\mu\nu} (k)\right] A_\nu (k) 
	= \mathcal{J}^\mu (k),
	\end{eqnarray}
which is equivalent to the photon equation of motion (\ref{phot_EOM}) but in momentum space.

Thus the photon polarization tensor which contains contributions from charged bosons
and fermions, $\Pi_{\mu\nu} (k) =
	\Pi_{\mu\nu}^{B} (k) + \Pi_{\mu\nu}^{F} (k)$,
and the electromagnetic current 
$\mathcal{J}_{\mu} $ involved in eq. (\ref{Phot_EOM_Pi_munu}) are explicitly found 
in the lowest order: 
	\begin{eqnarray}
	\label{phot_pol_tensor_bos}
	\Pi_{\mu\nu}^{B} (k) 
	= e^2 \hspace{-0.1cm} \int \frac{d^3q}{(2 \pi)^3 E} 
	\left[ f_B (E) + \bar f_B (E) \right] 
	\left[  \frac{1}{2} \, \frac{(2q - k)_\mu (2q - k)_\nu}{(q - k)^2 - m_B^2}
	+\frac{1}{2} \, \frac{(2q + k)_\mu (2q + k)_\nu}{(q + k)^2 - m_B^2} 
 -   g_{\mu\nu}\right],
	\end{eqnarray}
	\begin{eqnarray}
	\label{phot_pol_tensor_fer}
	\Pi_{\mu\nu}^{F} (k)
	\hspace{-0.1cm}  &= \hspace{-0.1cm} & 2 e^2 \int \frac{d^3q}{(2 \pi)^3 E} 
	\left[ f_F (E) + \bar f_F (E) \right] \times
	\cr\cr 
	&& \left[
	\frac{q_\nu (k+q)_\mu - q^\rho k_\rho g_{\mu\nu} + q_\mu (k+q)_\nu}{(k+q)^2-m_F^2} 
	+ \frac{q_\nu (q-k)_\mu + q^\rho k_\rho g_{\mu\nu} + q_\mu (q-k)_\nu}{(k-q)^2-m_F^2}
	\right],
	\end{eqnarray}
	\begin{eqnarray}
	\label{current}
	\mathcal{J}_{\mu} &=& - e \int \frac{d^4x}{(2\pi)^4}\exp^{-ikx} 
	\int \frac{d^3q}{(2 \pi)^3} \frac{q_{\mu}}{E} 
	\bigg[ f_B(E) - \bar f_B(E) 
	- 2 \bigg( f_F (E) - \bar f_F (E)  \bigg) \bigg]. 
	\end{eqnarray}
In eqs. (\ref{phot_pol_tensor_bos}) - (\ref{current}) $k^{\mu} \equiv (\omega, {\bf k})$ and 
$q^{\mu} \equiv (E, {\bf q})$ are 
respectively the photon and the scalar/spinor four momenta, $f_q$ and $f_{\bar q}$ are the 
particle (antiparticle) distribution functions and $g_{\mu\nu} = (+---)$.
We assume the following charge convention: the bosons have electric charge $+e$, 
while fermions 
have electric charge $-e$. Of course the charge of antiparticles
has the opposite sign.
It is worth to stress that the total $\Pi^{\mu\nu}$ as well as its bosonic and fermionic 
components separately satisfy the transversality condition $k^\mu \Pi_{\mu\nu} = 0$.
The last term in eq. (\ref{phot_pol_tensor_bos}) describes the contribution from the 
tadpole diagram and coincides with that found in ref.~\cite{leupold}. 
Evidently without this term the tranversality condition would be violated.

In what follows we use $\Pi^{\mu\nu}$ to derive plasma frequency, 
Debye screening mass, and dispersion relation for the  
photon propagation in plasma.

\section{Photon propagation in plasma}
\label{sec_coll_phen}

An electric field of a test particle is known to be screened by the plasma polarization. 
In the static case i.e. for $\omega =0$ it follows from eq. (\ref{Phot_EOM_Pi_munu}): 
	\begin{eqnarray}
	\label{photon-eq-00}
	(|\mathbf{k}|^2 - \Pi_{00}) A_0 = - q,
	\end{eqnarray}
where $q$ is a (small) charge of the test particle.
When $\Pi_{00}$ is independent on the photon momentum $\mathbf{k}$, 
eq. (\ref{photon-eq-00}) becomes the usual equation for a scalar field 
with mass $m = \sqrt{- \Pi_{00}}$. 
In this case which is usually realized at least for small $|{\bf k}| $
the Debye mass $m_D$ is equal to $\sqrt{- \Pi_{00}}$ and the 
electrostatic potential turns from the Coulomb to the Yukawa one:
	\begin{eqnarray}
	\label{yukawa}
	U = \frac{q}{4\pi}\,\frac{\exp(-m_D r)}{r}.
	\end{eqnarray}
Hence the Debye mass coincides with the inverse of the screening length. 
More generally the Debye mass is defined as the position of the poles of the
inverse operator $(|\mathbf{k}|^2 - \Pi_{00})^{-1}$ and in presence of the Bose
condensate the simple relation $m_D = \sqrt{- \Pi_{00}(\omega = 0)} $ becomes 
invalid  (see below).

There may be some ambiguities in definition of the screening potential
at higher orders in electric charge, $e^4$ and higher, and
in non-Abelian gauge theories where the result may violate gauge
invariance, as discussed in ref. ~\cite{rebhan-93}.
However, we consider here only Abelian $U(1)$ theory and in the
lowest, $e^2$ order. Moreover, as one can see below,
we have calculated the asymptotics of  the screening potential at
large distance in a self-consistent way by the location of the singularities
in the complex $k$--plane, which are situated at non-zero $k$.

Another quantity of interest is the plasma frequency $\omega_p$, which enters 
the dispersion relation of electromagnetic waves propagating in plasma. It is
defined as the limit of ${\bf k} \rightarrow 0 $ of certain space-space components 
$\Pi_{ij}$. Using the transversality condition, $k^\mu \Pi_{\mu\nu} = 0$,
we can decompose the photon polarization operator in a medium in terms of
two scalar functions, $a(\omega, |{\bf k}|)$ and $b(\omega, |{\bf k}|)$:
\begin{eqnarray}
\Pi_{ij} = a\left(\delta_{ij} - \frac{k_i k_j}{{\bf k}^2}\right)
+ b\frac{k_i k_j}{{\bf k}^2},\,\,\,\,
\Pi_{0j} = \frac{k_j}{\omega}\, b,\,\,\,\,
\Pi_{00} = \frac{ {\bf k^2}}{\omega^2} \, b.
\label{Pi-decomp}
\end{eqnarray}
We have assumed here that the medium is isotropic and so $a$ and $b$ depend only on the
absolute value of the photon momentum.

One can check from the presented above
explicit expressions (\ref{phot_pol_tensor_bos}) and (\ref{phot_pol_tensor_fer}) that $a = b$ in the limit ${\bf k} =0$.
In other words, in this limit $\Pi_{ij} \sim \delta_{ij}$. The plasma frequency is
determined by the equation:
\begin{eqnarray}
\omega_p^2 = b(\omega, |{\bf k }= 0|) = a(\omega, |{\bf k }= 0|).
\label{omega-p}
\end{eqnarray}

Since the distribution functions depend only upon the energy, the integral over
angles in eqs.  (\ref{phot_pol_tensor_bos}) and (\ref{phot_pol_tensor_fer}) can be taken. In particular for time-time
component of the polarization tensor in the limit of $\omega = 0$ we find:
\begin{eqnarray}
\Pi_{00}^B = - \frac{e^2}{2\pi^2}\,\int_0^\infty\,\frac{dq q^2}{E}
(f_B +\bar f_B)\left( 1 +\frac{E^2}{k q}\,\ln\bigg|\frac{2q +k}{2q -k}\bigg| \right),
\label{Pi-bos-int} \\
\Pi_{00}^F = - \frac{e^2}{\pi^2}\,\int_0^\infty\,\frac{dq q^2}{E}
(f_F +\bar f_F)\left( 1 +\frac{E^2}{k q}\,\ln\bigg|\frac{2q +k}{2q -k}\bigg| \right).
\label{Pi-ferm-int} 
\end{eqnarray}
Here and in what follows $k$ and $q$ are respectively the absolute values of the spatial 
component of the photon and charged particle momenta.
The argument of the logarithm comes in absolute value because 
the Green's functions in the perturbative expansion appear in the
combinations $G(q+k) + G^* (q-k)$. 
So the imaginary part of $\Pi_{00}$ in the considered limit vanishes.

It is even simpler to calculate the space-space components, $\Pi_{ij}$ in the
limit of zero  photon momentum, $k = 0$:
\begin{eqnarray}
\Pi_{ij}^B &=& \frac{e^2}{2\pi^2}\,\delta_{ij}\,\int \frac{dq q^2}{E}
\left(1 - \frac{4}{3}\,\frac{q^2}{4E^2 - \omega^2}\right) (f_B+\bar f_B ),
\label{Pi-ij-B} \\
\Pi_{ij}^F &=& \frac{4e^2}{\pi^2}\,\delta_{ij}\,\int \frac{dq q^2}{E}
\frac{E^2 - q^2/3}{4E^2 - \omega^2}\, (f_F+\bar f_F ).
\label{Pi-ij-F} 
\end{eqnarray}
It is clear from these expressions that in the limit $k=0$ functions 
$a$ and $b$ (\ref{Pi-decomp}) are equal. 

At $\omega = 2m$ the polarization operator acquires a non-zero imaginary
part which corresponds to the threshold of two charged particles production
by the photon. For massless charged particles the threshold is at $\omega =0$
and one has to take into account that the effective photon mass in plasma is 
non-zero (it is essentially the plasma frequency). This can  be done using
resummation technique. Accordingly the position of the threshold moves a little,
as $\sim e T$. This is not essential for our consideration. Moreover electrodynamics
with massless charged particles has serious infrared problems.

Let us now apply these results for the calculation of the plasma frequency and the 
Debye mass in some special cases which have been considered in the literature.

\emph{Massless SpQED and SQED with vanishing chemical potentials and high temperature, 
$T\gg m_{B,F}$}:
In thermal equilibrium the distributions of bosons and fermions and their antiparticles
with zero chemical potentials 
have the usual Bose-Einstein or Fermi-Dirac form:
	\begin{eqnarray}
	\label{No_Bose_defs}
	f_B(E,T) &=& \bar f_{B} (E,T) = \frac{1}{\exp (E/T) - 1},
	\cr\cr
	f_F(E,T) &=& \bar f_{F} (E,T) = \frac{1}{\exp (E/T) + 1},
	\end{eqnarray}
where in high temperature limit we can neglect the particle mass i.e. we can assume 
$E = |\bf {q}|$. In this special case $\Pi_{00}$ does not depend on the photon 
momentum $\mathbf{k}$ and so $m_D = \sqrt{-\Pi_{00}}$. The integrals in 
eqs. (\ref{phot_pol_tensor_bos}) and (\ref{phot_pol_tensor_fer}) can be easily taken and we find
for the contributions from bosons and fermions respectively:
	\begin{eqnarray}
	\label{m-D-B-F}
	m_{D \, B}^2 ( m_B = 0) &=& m_{D \, F}^2 ( m_F = 0) = 	\frac{ 1 }{3} \, e^2 T^2,
	\cr\cr
	\omega_{P \, B}^2 ( m_B = 0) &=& \omega_{P \, F}^2 ( m_F = 0) = 
	\frac{ 1 }{9} \, e^2 T^2.
	\end{eqnarray}
Hence the total Debye mass and the plasma frequency for the Lagrangian (\ref{L_SQED}) are
	\begin{eqnarray}
	\label{m-D-B-massless}
	m_{D}^2 ( m_B = m_F = 0 ) = \frac{ 2 }{3} \, e^2 T^2, 
\label{m_d-tot} \\
	\omega_P^2 ( m_B = m_F = 0) =
	\frac{ 2 }{9} \, e^2 T^2.
\label{omega-p-tot}	
\end{eqnarray}
These expressions coincide with the published results in the lowest order
in the electromagnetic coupling, $e^2$, see e.g. \cite{Kraemmer:1994az,Kapusta:1989tk}. 

In the case of relativistic fermions with non-zero chemical potential $\mu$ we
obtain:
\begin{eqnarray}
m_D^2 (m_F = 0) = e^2\left( \frac{T^2}{3} + \frac{\mu^2}{\pi^2} \right).
\label{mD-of-T-mu}
\end{eqnarray}
This is the result found in ref.~\cite{fradkin}. 

\emph{Massive SpQED with non-zero chemical potential and low temperature, $T \rightarrow 0$}.
In thermal equilibrium the distributions of fermions and their antiparticles are in 
this case the Boltzmann distributions:
	\begin{eqnarray}
	\label{Boltz_distr}
	f_F(E,\mu,T) = e^{(\mu - E)/T},\,\,\,\,
\bar f_{F} (E,\mu,T) = e^{-(\mu+E)/T} 
	\end{eqnarray}
and once again $\Pi_{00}$ does not depend on $k$, so it coincides with the Debye mass
squared. 
Hence in the case of relatively small chemical potentials, $\mu < m$, we obtain
	\begin{eqnarray}
	\label{Pi_00_T_fer_0}
	\Pi_{00} = \frac{4e^2}{T} \left( \frac{mT}{2\pi} \right)^{3/2} \,
e^{-m/T}\,\cosh (\mu/T).
	\end{eqnarray}
In the case of strongly degenerate,  $\mu\geq m$, nonrelativistic fermionic plasma, 
the contribution of anti-fermions may be neglected, while the fermion distribution
function has the form:
\begin{eqnarray}
f_F = \exp \left(\frac{\mu}{T} - \frac{q^2}{2m_F T} \right).
\label{f-F-nonrel}
\end{eqnarray}   
The chemical potential can be expressed through the number density of the fermions:
\begin{eqnarray}
n_F = \frac{\exp (\mu/T)}{\pi^2}\,\int dq q^2 e^{-q^2/2m_FT} .
\label{n-F}
\end{eqnarray}
There is a factor 2 in the above expression which counts two spin degrees of freedom.

Correspondingly the Debye screening mass for nonrelativistic fermions is 
\begin{eqnarray}
m_D^2 = \frac{e^2 n_F}{T},
\label{m-D-nonrel}
\end{eqnarray}
which coincides with the classical result, see e.g. book~\cite{landau-9}.

Analogously we find from eq. (\ref{Pi-ij-F}) the plasma frequency for nonrelativistic 
fermions:
\begin{eqnarray}
\omega^2_p = \frac{ e^2 n_F}{m_F},
\label{omega-p-nonrel}
\end{eqnarray}
which is also the classical result.

We have done these simple exercises to check the
validity of our results for $\Pi_{\mu\nu}$
comparing it to the known cases. Now we will turn to the calculations of the photon
polarization tensor in the medium with Bose condensate of charged scalars, which to our
knowledge was never done before.

\subsection{Bose-Einstein condensate}

If the density of particles is not equal to the density of antiparticles, the equilibrium
distributions are modified by introduction of chemical potentials. The distribution
functions take the form:
\begin{eqnarray}
f_{B,F} (E,\mu, T) = \frac{1}{\exp [ (E-\mu) / T ] \pm 1},
\label{f-of-mu}
\end{eqnarray}
where the plus sign stands for fermions and minus is for bosons. In equilibrium the chemical
potentials of particles and antiparticles have the opposite signs, $\bar\mu = -\mu$.
Evidently chemical potential of bosons is bounded from above, $\mu_B \leq m_B$.

If the asymmetry between bosons and anti-bosons reaches so large value that $\mu_B =m_B$
is not sufficient to ensure such a large asymmetry, Bose condensate would be formed and
the distribution functions take the form:
\begin{eqnarray}
 \label{f-B-cond}
	f_B^{(C)} (E,C,T)&=& \frac{1}{\exp [ (E-m_B) / T ] - 1} + C \, \delta({\bf q})
\equiv f_B(E,m_B,T) +  C \, \delta({\bf q}),\\
\label{bar-f-B}
\bar f_B (E,-m_B,T) &=& \frac{1}{\exp[ (E+m_B) / T ]  - 1},
\end{eqnarray}
where a new constant parameter $C$ is the amplitude of the condensate. 

The distributions in eq. (\ref{f-B-cond})
are the stationary solutions of the kinetic equation if and only if $\mu_B = m_B$
(or $\mu_{\bar B} = m_B$). Distribution (\ref{f-B-cond}) with $C\neq 0$ describes non-vanishing
number density of bosons in a vanishingly small momentum interval near $q = 0$. In this sense 
it is a classical field configuration $\phi (t) = \phi_0 \exp (im_Bt)$, where the
field amplitude is related to the condensate amplitude as $2 \phi^2_0 m = C / (2\pi)^3$. On the other
hand, it evidently can be interpreted as a collection of particles at rest.
 
Once the boson charge density $\mathcal{J}^B_0$ is fixed, it is possible to calculate the critical temperature $T_C$ when the phase transition takes place and the Bose condensate is formed.
This temperature can be calculated from eq. (\ref{current}) imposing
	\begin{eqnarray}
	\label{T-c-calc}
	\mathcal{J}^B_0 (\mu_B = m_B, C=0) = \mathcal{J}^B_0 (T_C),
	\end{eqnarray}
from which follows the equation
	\begin{eqnarray}
	\label{T-c}
	T_C = \sqrt{\frac{3\mathcal{J}^B_0}{e \, m_B}},
	\end{eqnarray}
which coincides with what we have found in the literature - 
see e.g. ref.~\cite{Kapusta:1981aa}.

We can express the amplitude of the condensate through the electric charge
density of the plasma. 
We impose the condition that the total electric charge density of fermions and bosons 
is zero. The condition of vanishing electric charge density may be not fulfilled e.g. 
in cosmology~\cite{charged-cosm}, but we postpone this case for future consideration. 
So we assume that $\mathcal{J}_0 = 0$ in eq. (\ref{current}). Expressed through
the particle occupation numbers $f_{B,F}$ this condition reads:
	\begin{eqnarray}
	\label{El_charge_zero}
	\int d^3q \,  \left[ f_B - \bar f_B - 2(f_F -  2\bar f_F)  \right] &=& 0.
	\end{eqnarray}
Here we skipped the arguments of $f_{F,B}$, which are explicitly
given in eqs. (\ref{f-of-mu}-\ref{bar-f-B}). The factor 2 in front of the fermionic 
distribution
functions is related to two spin degrees of freedom.

Hence the amplitude of the Bose condensate $C$ for
globally neutral plasma is equal to:
	\begin{eqnarray}
	C= - 4 \pi
	\int dq \, q^2
	 \left[  f_B(E_B,m_B,T) - \bar f_B(E_B,-m_B,T)
	 - 2 f_F(E_F,\mu_F,T) + 2 \bar f_F(E_F,\mu_F,T)
	\right].
	\end{eqnarray}
Here $E_{B,F} = \sqrt{ m_{B,F}^2 + q^2}$. The magnitude of the fermionic chemical potential,
$\mu_F$, is determined by the charge density of fermions. In the model under 
consideration it 
is a free parameter. We can imagine another more realistic model, when the bosons have the 
Yukawa coupling to electrons and neutrinos, $g \phi \bar\psi_\nu \psi_e $. This coupling 
leads to the process $\phi^+ \leftrightarrow \nu+e^+$. 
If the lepton number density is sufficiently large, chemical
potential of $\phi$ may reach the limiting value, $\mu_B = m_B$, and $\phi$-bosons would 
condense.

Now we can use eqs. (\ref{phot_pol_tensor_bos}), (\ref{Pi-bos-int}) and (\ref{Pi-ferm-int})
to calculate 
$\Pi_{00} (\omega =0,  k 
)$ which is needed to work out the 
electrostatic potential of a test charge in the plasma with Bose condensate:
	\begin{eqnarray}
	\label{Gen_P00_2}
	\Pi_{00}  &&  \!\!\!\!\!\!\!\!\!\!\!\!\!\! (\omega =0,  {k}  
	) =
	\frac{e^2}{(2\pi)^3} \, \frac{C}{m_B}
	\left( 1 + \frac{4m_B^2}{k^2} \right) \cr\cr
	&+&
	\frac{e^2}{2\pi^2} \int_0^\infty  \frac{dq \,q^2}{E_B} 
\left[f_B(E_B,m_B,T)+ \bar f_B(E_B,-m_B,T)\right] 	
	\left[1+\frac{ E^2_B}{ k q} \ln \bigg| \frac{2q +k}{2q - k} \bigg| \right] \cr\cr
	&+& 
	\frac{e^2}{2\pi^2} \int_0^\infty \frac{dq\,q^2}{E_F} 
\left[f_F(E_F,\mu_F,T)+ \bar f_F(E_F,-\mu_F,T)\right] 	
	\left[2+\frac{ (4E_F^2 - k^2)}{2 k q} \ln \bigg| \frac{2q +k}{2q - k} \bigg| \right].
	\end{eqnarray}

It is rather surprising that the first term which appears only in the presence of the Bose
condensate has a singularity, $1/k^2$. However this is not the only singular in $k$ term.
There is also a singularity $1/k$ arising from the bosonic (but not anti-bosonic) 
contribution to $\Pi_{00}$ if $\mu = m_B$
even in the absence of condensate, i.e. for $C=0$, which appears from the integration
region
where $q$ is smaller or comparable to $k$. For small $q$ the distribution function is 
infrared singular,
\begin{eqnarray}
f_B (E_B,m_B,T) \approx 2m_B T/ q^2.
\label{f-B-small-q}
\end{eqnarray}
Usually this singularity is not dangerous because it is canceled by the integration 
measure, $\sim q^2$.
However, there is another factor to worry about. The logarithmic term behaves as
$k/q$ for $q>k$ and as $q/k$ for $q<k$. So the integral is finite but has $1/k$ singularity.
Indeed we can separate the integral into two parts $0<q<k/2$ and $k/2 <q< \infty$. 
It is convenient to introduce for the first part the new integration variable 
$x=2q/k$, so $0<x<1$. For the second part we introduce $y = k/2q$, so $y$ runs in the
same limits, $0<y<1$. In the limit of small $k$ we can expand $E_B \approx m_B + k^2x^2/8m_B$.
Correspondingly:
\begin{eqnarray}
\exp\left[\frac{E_B-m_B}{T}\right] -1 \approx k^2 x^2 /8m_B T.
\label{exp-expand}
\end{eqnarray}
So the integral is reduced to
\begin{eqnarray}
\int_0^1 \frac{dx}{x} \ln \left( \frac{1+x}{1-x}\right) = \frac{\pi^2}{4}.
\label{int}
\end{eqnarray}
The same contribution comes from the second part of the integral. So finally
we obtain for the singular in $k$ part of the photon polarization tensor
in the case of $\mu =m_B$ and $C=0$: 
\begin{eqnarray}
\Pi_{00} = \frac{e^2 m_B^2 T}{2k}.
\label{1-over-k}
\end{eqnarray}
Numerical calculations without expansion of the energy and the exponent gives 
a very close result.


Thus for zero frequency and in the limit of small $k$ the time-time component of 
the polarization tensor has the form:
\begin{eqnarray}
\Pi_{00} =  e^2\left[ \frac{2T^2}{3} + \frac{m_B^2 T}{2k} +
\frac{1}{(2\pi)^3} \, \frac{C}{m_B}
	\left( 1 +  \frac{4m_B^2}{ k^2} \right)  \right].
\label{final-Pi-00}
\end{eqnarray}
This form of $\Pi_{00}$ determines the position of the singularity points 
in the complex k-plane at low $k$, which in turn governs the asymptotics 
of the screening potential at large $r$.
For simplicity we took here the high temperature case, $T\gg m_{F,B}$ when the
k-independent terms have a very simple form but the analysis can be done for any $T$.
The general form of $\Pi_{00}$ is presented in eq. (\ref{max-eq}). According to the result
above:
\begin{eqnarray}
m_0^2 &=& {2T^2}/{3} + C/{[(2\pi)^3m_B]} \nonumber\\
m_1^3 &=&  {m_B^2 T}/{2} \nonumber \\
m_2^4 &=& {4Cm_B}/{(2\pi)^3}  .
\label{m-0-1-2}
\end{eqnarray}

To determine the Debye screened potential we have to make the Fourier transformation of the
inverse of the photon equation of motion $k^2 - \Pi_{00} = 0$:
\begin{eqnarray}
U(r) = q \int \frac{d^3 k}{(2\pi)^3} \frac{\exp (i {\bf k r)}}{k^2 - \Pi_{00} (k)} =
\frac{q}{2\pi^2}\int_0^\infty \frac{dk k^2}{k^2 - \Pi_{00} (k)}\,\frac{\sin kr}{kr}.
\label{U-of-r}
\end{eqnarray}
If $\Pi_{00}$ is an even function of $k$, as it is usually the case, the integration over
$k$ can be extended to the interval from $-\infty$ to $+\infty$ and the integral can be
taken as a sum over residues of the integrand. For example, if the term proportional to
$1/k$ can be neglected (low temperature case), 
$\Pi_{00}$ is evidently even and its poles can be easily found:
\begin{eqnarray}
k_j^2 = -\frac{e^2 m_0^2}{2}\pm \sqrt{\left[\frac{e^4 m_0^4}{4} - e^2m_2^4 \right]}
\approx \pm i e m_2^2.
\label{k-j-2}
\end{eqnarray}
The last approximate equality is formally true in the limit of small $e$.
 
Thus in presence of Bose condensate
the ``Debye'' poles have non-zero real part:
\begin{eqnarray}
k_j = \pm \sqrt{e} m_2 \exp (\pm i\pi/4) \equiv
k_j' + ik''_j,
\label{k-j}
\end{eqnarray}
while $k'=0$ in the usual case. Non-zero $k'$ leads
to the oscillating behaviour of the potential
\begin{eqnarray}
U(r)_j \sim q\, \frac{\exp ( - \sqrt{e/2} m_2 r) \cos (\sqrt{e/2} m_2 r)}{r}.
\label{U-j}
\end{eqnarray}

If the term proportional to $1/k$ is present in $\Pi_{00}$
the calculations of the potential are slightly more complicated. Now the integration
path in the complex $k$ plane cannot be extended to $-\infty$ but the integration 
should be done along the real axis from 0 to $\infty$, then along infinitely large 
quarter-circle, and along the imaginary axis from $-\infty$ to 0. The result would
contain the usual contribution from the poles in complex $k$-plane and the integral
over imaginary axis. The former gives the usual exponentially decreasing
potential, while the latter gives a power law decrease:
\begin{eqnarray}
 U \sim q\, m_1^3/(e^2 m_2^8 r^6).
\label{power-U}
\end{eqnarray}
Notice that the potential is inversely proportional to the electric charge squared.

These strange results possibly indicate that the approximation of the uniform density
plasma is not accurate near the test charge in the presence of Bose condensate and
the polarization of the medium is very strong, which is rather natural for particles
at rest.

It is worth to notice that Debye screening is non-analytic in $e$, i.e. 
$m_D \sim e^{1/2}$.

On the other hand, there is no anomalous behavior for the propagating modes,
i.e. for $\Pi_{ij}$. In particular the plasma frequency in the presence of the 
condensate acquires simply a constant contribution 
$\delta \omega_p^2 = e^2 C /(2\pi)^2 m_B $.

\section{Calculations in another gauge}
\label{Calcs-in-other-gauge}

Though our result is gauge invariant and for the calculation of the Debye
screening length we do not need to fix the gauge, still it may be instructive to make 
the calculations in the gauge used in ref.~\cite{Linde:1976kh}. The homogeneous state
of charged scalars with non-zero charge density was described in this paper as
\begin{eqnarray}
&&\phi=\phi_0 = const, \label{phi-adl}\\
&&A_\mu^{(0)}=(m_B/e)\, \delta_\mu^0.
\label{A-mu-adl}
\end{eqnarray}
It can be easily seen that this solution is a gauge transform of the zeroth 
order state (in coupling $e$) used in the present paper:
\begin{eqnarray}
&&\phi = \phi_0 \exp (i m_B t), \label{phi-dlp} \\
&&A_\mu = 0.
\label{A-mu-dlp}
\end{eqnarray}
Evidently such a state of $\phi$ describes a collection of bosons at rest i.e.
of a Bose condensate. The electric charge density of the condensate can be read off
equation (\ref{EOM_current}) and is equal to:
\begin{eqnarray}
J_0^{(C)} = 2e m_B |\phi_0|^2.
\label{J-C}
\end{eqnarray}
Expressions (\ref{phi-adl},\ref{A-mu-adl}) 
and (\ref{phi-dlp},\ref{A-mu-dlp}) lead to the same  
result for the electric charge density. Comparing it with the charge density 
described by the equilibrium distribution (\ref{f-B-cond}) we find that we have
to identify:
\begin{eqnarray}
C/(2\pi)^3 = 2 m_B |\phi_0|^2.
\label{C}
\end{eqnarray}

Perturbation theory is less convenient in gauge (\ref{phi-adl},\ref{A-mu-adl})
because of large value of the background potential $A_0^{(0)} \sim 1/e$. We need to 
make the expansion:
\begin{eqnarray}
A_\mu \rightarrow \frac{m_B}{e} \,\delta_\mu^0 + A_\mu,
\label{A-1}\\
\phi = \phi_0 + \phi_{q},
\label{phi-1}
\end{eqnarray} 
where $A_{\mu}$ is the potential of the physical electromagnetic field in the
plasma. It may describe e.g. Coulomb-like field around test charge when we discuss
Debye screening or propagating waves in plasma when we talk about plasma frequency.
This $A_\mu$ is equal to $A_\mu$ considered above when we worked in the gauge
defined by eqs. (\ref{phi-dlp},\ref{A-mu-dlp}). The quantum deviation 
from the condensed state of the scalar field, $\phi_q$, is supposed to be zero
on the average. Moreover, we assume for simplicity that the plasma temperature
is zero and thus $\langle \phi_q^2\rangle$ vanishes as well (if the vacuum quantum
fluctuations are subtracted). So $\phi$ enters only into description of
virtual particles through the Green's function.

The equation of motion of $\phi_q$ has the form:
\begin{eqnarray}
(\partial^2 - 2im_B\partial_0) \phi_q = 2e m_B  A_0 \phi_0 +
e^2 (A_\mu)^2 (\phi_0 + \phi_q )
+ ie \left[2A^\mu \partial_\mu + (\partial_\mu A^\mu)\right] \phi_q.
\label{d2-phi-1}
\end{eqnarray}
Only the first term in the r.h.s. of this equation will be essential in what follows.

It is straightforward to write down the equation for $A_\mu$:
\begin{eqnarray}
\partial^2 A_\mu - \partial_{\mu} \left(\partial^\nu A_\nu\right) = 
J_\mu^{(F)} - 2e m_B\delta_\mu^0 \left[\phi_0^2 - 2\phi_0 {{R}}e \phi_q\right] + ... \;.
\label{d2-A1}
\end{eqnarray} 
Here we retained only terms which are essential for the calculation of the condensate
impact on the photon propagation in plasma. The missing terms can be easily found from
equation (\ref{partial_drv_Fmunu}). 
The first term in eq. (\ref{d2-A1}) is the fermionic current. Together 
with the second term they make the total current in the plasma, which we assume as
above to be zero, $J_\mu^{(F)} - 2e m_B\delta_\mu^0 \phi_0^2 =0$. The remaining term 
in the r.h.s. can be 
found by perturbative solution of  eq. (\ref{d2-phi-1}):
\begin{eqnarray}
\phi_q = 2em_B \int G(x-y) \left[ \phi_0 A_0 (y) + ... \right].
\label{phi-q}
\end{eqnarray}
As above, we make the Fourier transformation (\ref{A-mu-k})
and obtain the correction to the time-time component of the polarization operator 
from the field $\phi_0$:
\begin{eqnarray}
\delta \Pi_{00} = e^2 m_B^2 \phi_0^2 / {\bf k}^2.
\label{delta-Pi-00}
\end{eqnarray}
Keeping in mind identification (\ref{C}) we find that it is exactly the same result
which we have found above working in terms of equilibrium distribution with Bose
condensate. This finalizes the argument that both descriptions (in both gauges) are
equivalent and both states $\phi= \phi_0 \exp (imt) $ and $A_0 = 0$ and 
$\phi=\phi_0$ and $A_0 = m_B/e$ describe the same collection of particles at rest.

\section{Conclusion}
\label{sec_concl}

We have calculated the polarization operator of photon in the plasma with charged
scalar bosons and spin one half fermions. The calculations have been done in the
lowest order in electromagnetic coupling constant, $e$. Our technique is slightly 
different from the standard one, that is from
either imaginary time method applicable to 
thermal equilibrium case or real time method which leads to the doubling of states.
We simply solved perturbatively the operator equations of motion for the 
charged fields $\Phi$. The solution in the lowest order
typically has the form
	\begin{eqnarray}
	\label{phi-1-x}
	\Phi_1 (x) = 
	\int dy G(x-y) \mathcal{J} (\Phi_0), 
	\end{eqnarray}
where $\Phi_{0,1}$ are the charged fields in zeroth and first approximation
in $e$ respectively and $\mathcal{J}$ is the current entering the equation of motion. 
This solution was substituted into the Maxwell equation for the
quantum electromagnetic field and the average of quantum operators $\Phi_0$  was taken
over the medium. In this way the effect of medium for arbitrary occupation numbers 
of the charged particles (not necessary equilibrium) can be taken into 
account. We checked our results comparing them with the known cases of Debye
screening and the plasma frequency at low and high $T$ and nonzero chemical
potentials.

The new part of the work consists of calculation of the polarization operator
in the presence of Bose condensate of charged scalar particles. Though it is
technically simple, there are no such calculations in the literature, as 
far as we know. Physically
such a case can be realized if there is a significant asymmetry between charged
fermions (electrons and positrons). Assuming that the global charge of the plasma
is zero (though it is not necessarily so) we can find the number density of
charged scalars in the condensate. To this end the charge asymmetry between fermions
must be sufficiently large so that the maximum allowed chemical potential of
bosons, $\mu_B = m_B$ is not enough to secure the vanishing of
total charge. The necessary
neutralization can be achieved by the charge density of the condensate,
aving the distribution $ f_c = C \delta ({\bf q})$. The corresponding correction to the 
polarization operator has a singularity, i.e.
a pole at zero three-momentum of photon, 
$ 4C e^2 m_B/ (2\pi)^3 k^2 +  e^2 m_B^2 T / 2k$.
Probably the origin of this singularity is a large mobility of particles with 
zero momentum under the influence of an external electric field. Energetic particles
are much less ``eager'' to screen the test charge. 

Due to these infrared singular terms the Debye screening length
becomes parametrically shorter, $\lambda_D \sim 1/\sqrt{e}$ (if $C\neq 0$), instead
of the standard one $\lambda_D \sim 1/e$ which is true for the plasma of charged fermions
and/or bosons with $\mu_B <m_B$. 
It is noteworthy that the Debye screening is
now a non-analytic function of the electromagnetic coupling $e$. 
The screened electrostatic potential in the presence of Bose condensate of charged scalars
has an oscillating behavior superimposed on the exponential decrease, eq.~(\ref{U-j}).  
The presence of $1/k$
term which appears at non-zero temperature when $\mu=m_B$, leads to weaker, power law, decrease
of the screened potential, which is inversely proportional to the electromagnetic coupling
constant squared, eq.~(\ref{power-U}).

There are no known charged stable scalar fields, so the considered situation
may be unrealistic. However, it is not excluded that an extension of the minimal
standard model will demand stable charged Bose fields. If this is the case, the
calculated photon dispersion relation may be of interest in cosmology or in
dense stellar environment. For example, we may expect condensation of di-quarks
in quark stars. On the other hand, stability of scalars may be irrelevant because
even unstable particles may condense in a dynamical equilibrium state.

Bose condensation of charged scalars can possibly be realized also in the following 
situation. Let there be high temperature plasma of $e^{\pm}$, $\nu$, and $\bar \nu$
with a considerable excess of $e^-$ over $e^+$ and of $\nu$ over $\bar \nu$, i.e.
there are large lepton and electric charge asymmetries. We assume that there are 
also charged pions (or some other scalar particles) in the plasma to ensure 
zero total electric charge. The
condensed scalars may be stabilized by large chemical potentials of 
neutrinos, such that the decay $\pi^+ \rightarrow e^+ \nu$ 
is not allowed because
the Fermi states with $E = m_\pi/2$ are already occupied. Such a state might 
possibly exist in exotic stars.

{\vspace{0.8cm} \noindent \bf Acknowledgments}
The authors would like to thank to L. Bonanno, A. Drago, G. Gabadadze, A. Linde, and 
M. Voloshin for stimulating discussions and criticism. 
G. P. acknowledges the kind support provided by the grant program PASPA-UNAM and PAPIIT-UNAM, number IN112308.


\end{document}